\documentclass[reprint,jkps,fleqn,showpacs,showkeys,superscriptaddress]{revtex4-2}
\usepackage[pdftex]{graphicx}
\usepackage{color}
\usepackage{float}
\usepackage{amssymb}
\usepackage{amsmath}
\usepackage{bm}
\usepackage{url}

\usepackage{breakurl}
\usepackage[breaklinks]{hyperref}

\begin{document}
\setcounter{page}{1}
\title[]{Effectiveness of vaccination and quarantine policies to curb the spread of COVID-19}
\author{Gyeong Hwan \surname{Jang}}
%\email{rud2612@gmail.com}
\affiliation{Department of Applied Physics, Hanyang University, Ansan 15588, Republic of Korea}
\author{Sung Jin \surname{Kim}}
\affiliation{Department of Urology, University of Ulsan College of Medicine, Ulsan 44610, Republic of Korea}
\author{Mi Jin \surname{Lee}}
\email{mijinlee@hanyang.ac.kr }
\affiliation{Department of Applied Physics, Hanyang University, Ansan 15588, Republic of Korea}
\author{Seung-Woo \surname{Son}}
\email{sonswoo@hanyang.ac.kr }
\affiliation{Department of Applied Physics, Hanyang University, Ansan 15588, Republic of Korea}
\affiliation{Department of Applied Artificial Intelligence, Center for Bionano Intelligence Education and Research, Hanyang University, Ansan 15588, Republic of Korea}
\date{\today}

\begin{abstract}
A pandemic, the worldwide spread of a disease, can threaten human being in the social as well as biological perspectives and paralyze existing living habits. To stave off the more devastating disaster and return to a normal life, people make tremendous efforts at multiscale levels from individual to worldwide: paying attention to hand hygiene, developing social policies such as wearing mask, social distancing, quarantine, and inventing vaccine and remedy. Regarding the current severe pandemic, namely the coronavirus disease 2019, we explore the spreading-suppression effect when adopting the aforementioned efforts. Especially the quarantine and vaccination are considered since they are a representative primary treatment for block of spreading and prevention at the government level. We establish a compartment model consisting of susceptible (S), vaccination (V), exposed (E), infected (I), quarantined (Q), and recovered (R) compartments, called SVEIQR model. We look into the infected cases in Seoul and consider three kinds of vaccines, Pfizer, Moderna, and AstraZeneca. The values of the relevant parameters are obtained from empirical data of Seoul and clinical data for vaccine and estimated by Bayesian inference. After confirming that our SVEIQR model is plausible, we test the various scenarios by adjusting the associated parameters with the quarantine and vaccination policies around the current values. The quantitative result obtained from our model could suggest a guideline for policy making on effective vaccination and social policies.
\end{abstract}
\keywords{Epidemic model, Vaccination, Quarantine}
\maketitle

\section{INTRODUCTION}
\label{sec:introduction}

%%% the most famous pandemic in the past
The Black Death, the most fatal pandemic in history, not only wiped out human lives but also made human suffer their economic, social, political, and religious upheaval across Europe in the fourteenth century~\cite{aberth2012environmental, RN53}. The continent-wide severe destruction caused by the pandemic has given the people at that time and their descendant the lesson on the importance of preventing the incidence of a disease itself as the primary treatment and its global spreading as the secondary one.
%%% the present situation about the epidemic spreading
Despite this lesson, humankind has continued to suffer the great and small outbreaks of epidemics so far. Ironically, the surprising and remarkable development of techniques and knowledge have helped the eradication and prevention of diseases but, meanwhile, used to provoke an unexpected spreading of new diseases, e.g., the creation of mutant plague in a laboratory and the fast transportation to be able to accelerate the spreading~\cite{Brockmann2013}. 

Until recently, some worldwide epidemics have led to social and medical tension, such as severe acute respiratory syndrome in 2002~\cite{sars}, Middle East respiratory syndrome in 2013~\cite{mers}, and the coronavirus disease 2019, abbreviated as COVID-19~\cite{covid}, which is the most severe pandemic in decades so far. While the outbreak sizes of the first two diseases are 8,096 and 2,449 respectively~\cite{sars, mers}, the COVID-19 have yielded about over 4 hundred million worldwide during the first two years from its incident~\cite{covid}.

%%% previous researches (compartmental model)
To understand its peculiar behavior of spreading as rarely encountered before, many researchers have attempted to examine the behavior in terms of social epidemiology as well as biology. The general and popular theoretical model is of the compartmental model where a population is imposed by a compartment with a label standing for its states, such as the susceptible-infected-recovered (SIR) model or the susceptible-exposed-infected-recovered (SEIR) model~\cite{SEIR, SEIR2, SEIR3, SEIR4}. The spreading of COVID-19 is also modeled as a range of variants of the SIR or SEIR model~\cite{SIR-SEIR, SEIR5Covid}.

The primary reason for modeling the epidemics is to predict the spreading pattern in the real world and then to aid in preventing and suppressing the spreading. However, the theoretical approaches allow us to understand minimal and essential factors of the spreading but are too limited to reflect complicated features in reality. To overcome the limits, the recent studies include the social policies, e.g., social distancing, banning group gatherings, and quarantine of the infected person~\cite{SEIQR, SEIQR2, SEIQR3, SIRX}. After inventing vaccines for the virus, in the modeling, the vaccination strategy has been also considered to efficiently utilize the vaccines within the limited quantities~\cite{SVEIR, SVEIR2, SVEIR3}. 

%%% our model
As primary external factors to suppress the epidemics, we focus on the \emph{quarantine} and the \emph{vaccination} strategy in the social and biochemical perspective, which are the direct action for obstructing the spread in their respective viewpoints under the current conditions. To grasp the effects of the quarantine and vaccination strategy, we propose a compartment model with considering the two factors as compartments, named as the susceptible-vaccinated-exposed-infected-quarantined-recovered (SVEIQR) model. It is a variant of the SEIR model rather than SIR model, to allow for a latent period of the virus represented as exposed (E) state where an individual is exposed to the virus but does not transmit it to others. This suggested model constructed as the compartment model still seems so simple to describe the reality in detail because it is hard to trace the networked structure among people in the outbreak. Nonetheless, we endeavor to give the reality to analysis as far as possible by deploying the empirical and clinical data and using the Bayesian inference to determine the relevant parameters.

%%% summary of results
The main finding is to suggest the quantitative result of the suppression effect under the current strategies or policies depending on the vaccine efficacy. The quantitative analysis allows us to predict the suppression effect when modifying the level of the strategies, compared to the present case, which policymakers may refer to.

%In this model, we reflect the following three novel features:
%i) %\emph{Realistic reproduction number} - 
%First, we calculate the transmission rate of COVID-19 from the empirical data using the Bayesian inference, and then utilize it in our SVEIQR model to reflect the actual spreading situation. Since the mathematical compartment model is suitable only for urban-sized simulation and well-mixed society, we apply the simulated model for Seoul, which is the largest city in South Korea and the second highly populated city in the world, using the infected case data of COVID-19.
%ii) %\emph{Clinical efficacy of vaccines} - 
%Second, utilizing the efficacy with confidence interval (CI) of three vaccines, Pfizer (95\%, CI: 90.0 -- 98.0\%), Moderna (94\%, CI: 88.0 -- 98.0\%), and AstraZeneca (70\%, CI: 56.0 -- 85.0\%), which are well known for their efficacy as clinical trials, we simulate to analyze the suppression effect of vaccination speed corresponding to the governmental policy in Seoul.  The vaccination speed corresponding to the current governmental policy of South Korea is about 70\% of the total population vaccinated for 9 months. We test the effect of vaccination speed for the three vaccines considering their efficacy. 
%iii) Third, comparing the vaccination speed and quarantine speed in our model, we tried to figure out the effects of each social policy. The total number of infected case is considered for the measurement of the impact.

This paper is organized as follows:
In Sec.~\ref{sec:model}, we present the SVEIQR model of infectious disease.
The Bayesian inference of the reproduction number is explained in Sec.~\ref{sec:reproduction_number}. The simulation results are displayed in Sec.~\ref{sec:results} followed by the summary and discussion.
%=========+=========+=========+=========+=========+=========+=========+=========

\begin{figure}
\includegraphics[width=0.85\columnwidth]{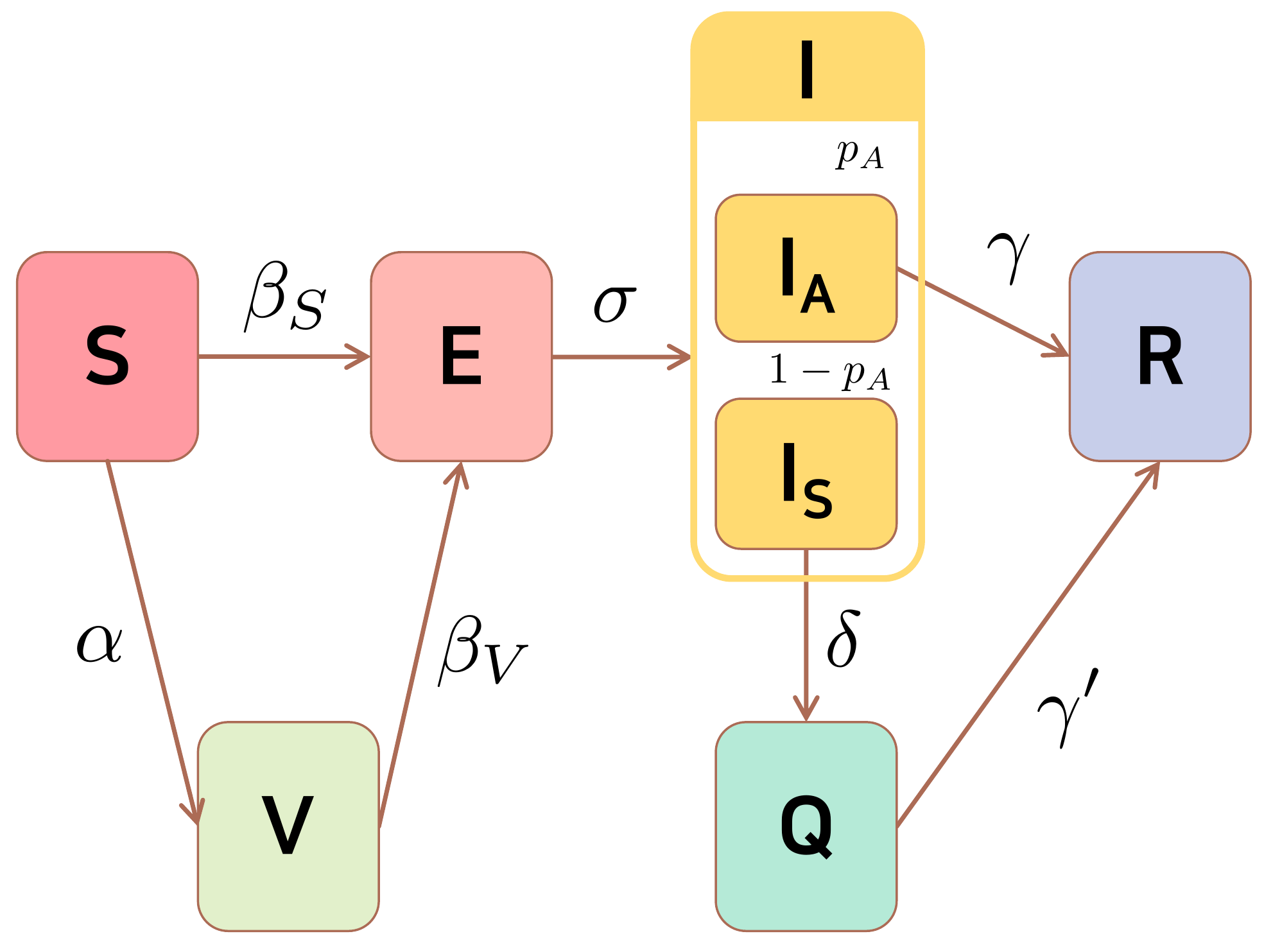}
\caption{An architecture of SVEIQR model. Susceptible (S), vaccinated (V), exposed (E), infected (I), quarantined (Q), and recovered (R) groups are considered. For the infected group, I, we divide into two subgroups which have symptom (${\rm I_S}$) and no symptom (${\rm I_A}$), respectively. There are two epidemic transmission rates $\beta_{S}$ and $\beta_{V}$. The transmission-after-vaccination rate $\beta_V$ is related with efficacy $\epsilon$ of vaccine and should be smaller than $\beta_S$. The inverse $\sigma$ describes the latent period of the disease. The $\gamma$ and $\gamma^{\prime}$ are the recovery rate. The vaccination speed $\alpha$ and quarantine speed $\delta$ represent the social strategy of how fast the government vaccinates the susceptible people and quarantines the infected people.}
\label{fig:model}
\end{figure}

\section{A compartmental model with %considering 
the quarantine and vaccination}
\label{sec:model}

%%% introducing our model, especially the three compartments E, V, Q.
To construct a realistic but minimal compartment model, we consider each representative factor including the feature of the virus itself, social policy, or prevention, leading to selecting the factors as incubation, quarantine, and vaccination. The first one is a biological feature (directly relevant to human) of the virus whose incubation period of the Delta strain is known as 4.1 days on average~\cite{who}, which means that symptoms after the first contact with the virus will appear by 4.1 days. We translate the incubation period as the transition rate of the E compartment to the infected compartment. In addition, one can also consider another biological features causing recovery and mortality from an infected state. The two state are definitely different but the same in such a way that they do not participate in spreading the infection. Hence, we herein only take account into the recovered (R) compartment in the rest of this paper.

Next, as a social policy, a government compels infected individuals as confirmed cases to go straight into quarantine (Q)~\cite{quarantine_who}. An infected individual will eventually become recovered, but one needs to distinguish the individual in the quarantined state and not because the infected and then consequently quarantined one (officially discovered by the government) does not propagate the disease.  To describe the course around the infected one in addition to the preexisting transition from I directly to R, we design the transition as I $\to$ Q $\to$ R by introducing the Q compartment. The transition rates from I to Q and from Q to R depend on the social infrastructure and policy associated with the inspection human power, the number of inspection stations, the number of the bed hospitals, and so on, while the direct transition I $\to$ R (recovery of undiscovered I) is relevant to natural healing.

The last factor, namely the vaccination (V) is an biochemical treatment right after the incident at the current situation in the absence of the disease remedy. The key target for vaccination basically focuses on the susceptible individual (although including the already infected ones), because its main purpose is to prevent the larger outbreak. %It depends on whether the vaccine uses whole-microbe approach (e.g., viral vector vaccine) or genetic approach (e.g., mRNA vaccine), 
Because the vaccine increases just the possibility not to catch a disease (dose not thoroughly block),  we consider that the vaccinated people can also turn into the exposed one to the disease. Thus, we describe the process associated with the vaccination as S $\to$ V $\to$ E. Also, note that the vaccination is an action associated with the combination of biochemical treatment (e.g., types of vaccines) and social policy and strategy (e.g., how to distribute vaccines).

To sum up, the compartments participating in our model are S, V, E, I, Q, and R, so we name this model SVEIQR model. We sketch the transition process between states in Fig.~\ref{fig:model}, and the governing rate-equations of the compartments are as follows: 
\begin{subequations}
\label{eq:rate_eqs}
    \begin{align}
        \frac{dS}{dt} &= -\beta_S \frac{SI}{N} - \alpha S \label{eq:rate_eq_s},\\
        \frac{dV}{dt} &= \alpha S - \beta_V \frac{VI}{N}\label{eq:rate_eq_v},\\
        \frac{dE}{dt} &= \beta_S \frac{SI}{N} + \beta_V \frac{VI}{N} - \sigma E\label{eq:rate_eq_e},\\
        \frac{dI}{dt} &= \sigma E - (1-p_A)\delta I - p_A \gamma I\label{eq:rate_eq_i},\\
        \frac{dQ}{dt} &= (1-p_A)\delta I - \gamma^{\prime} Q\label{eq:rate_eq_q},\\
        \frac{dR}{dt} &= \gamma^{\prime} Q + p_A \gamma I\label{eq:rate_eq_r},
    \end{align}
\end{subequations}
where each alphabet character of the compartment represents the number of the corresponding state. Provided the total population $N$ is conserved at any time, $S+V+E+I+Q+R=N$ or equivalently the sum of all above equations %in Eq.~(\ref{eq:rate_eqs}) 
should be zero. It is easier to refer to the schematic diagram in Fig.~\ref{fig:model} rather than Eq.~(\ref{eq:rate_eqs}) in order to understand the roles of several transition rates denoted by the Greek letters.

%%% description of parameters
The parameters $\beta_S$, $\sigma^{-1},$ and $\gamma^{-1}$ are respectively the contact/epidemic transmission rate, the average latent period for the disease, and the average infectious period (or the recovery probability from the disease per time), which are the same as those in conventional SEIR model. When considering the compartment V, we introduce two transition rates, i.e., the vaccination speed rate $\alpha$ and the transmission-after-vaccination rate $\beta_V$. To make sense, $\beta_V$ should be less than $\beta_S$ and also depends the performance of the vaccine, so we assume $\beta_V=(1-\epsilon)\beta_S$ with $\epsilon$ being the efficacy of the vaccine. This assumption is corroborated by the good accordance between the theoretical and experimental results. The previous research~\cite{KSKim} has modeled the inhibition effect as to the spreading of Borna disease virus with the same form $(1-\epsilon)\beta$ as ours. 

The compartment Q also requires two relevant transition rates. One is the quarantine rate $\delta$ and the inverse of the average quarantined period $\gamma^{\prime}$ (or equivalently release probability from quarantine). One can note that the quarantine individuals are also the infected ones. I and Q are biologically the same, and Q is the merely officially discovered I by the health authority. The health authority can take an action for quarantine only for the symptomatic infected people, not for the asymptomatic one, so we separate the infected individuals into the symptomatic and asymptomatic groups with the probability $1-p_A$ and $p_A$, respectively (i.e., $I_S = (1-p_A)I$ and ${I_A}=p_A{I}$). After the quarantine period, the released people from quarantine are regarded as being recovered from the disease, so they belong to the compartment R and we call the $\gamma^{\prime}$ the recovery rate as well. 

\begin{figure*}
\includegraphics[width=0.8\textwidth]{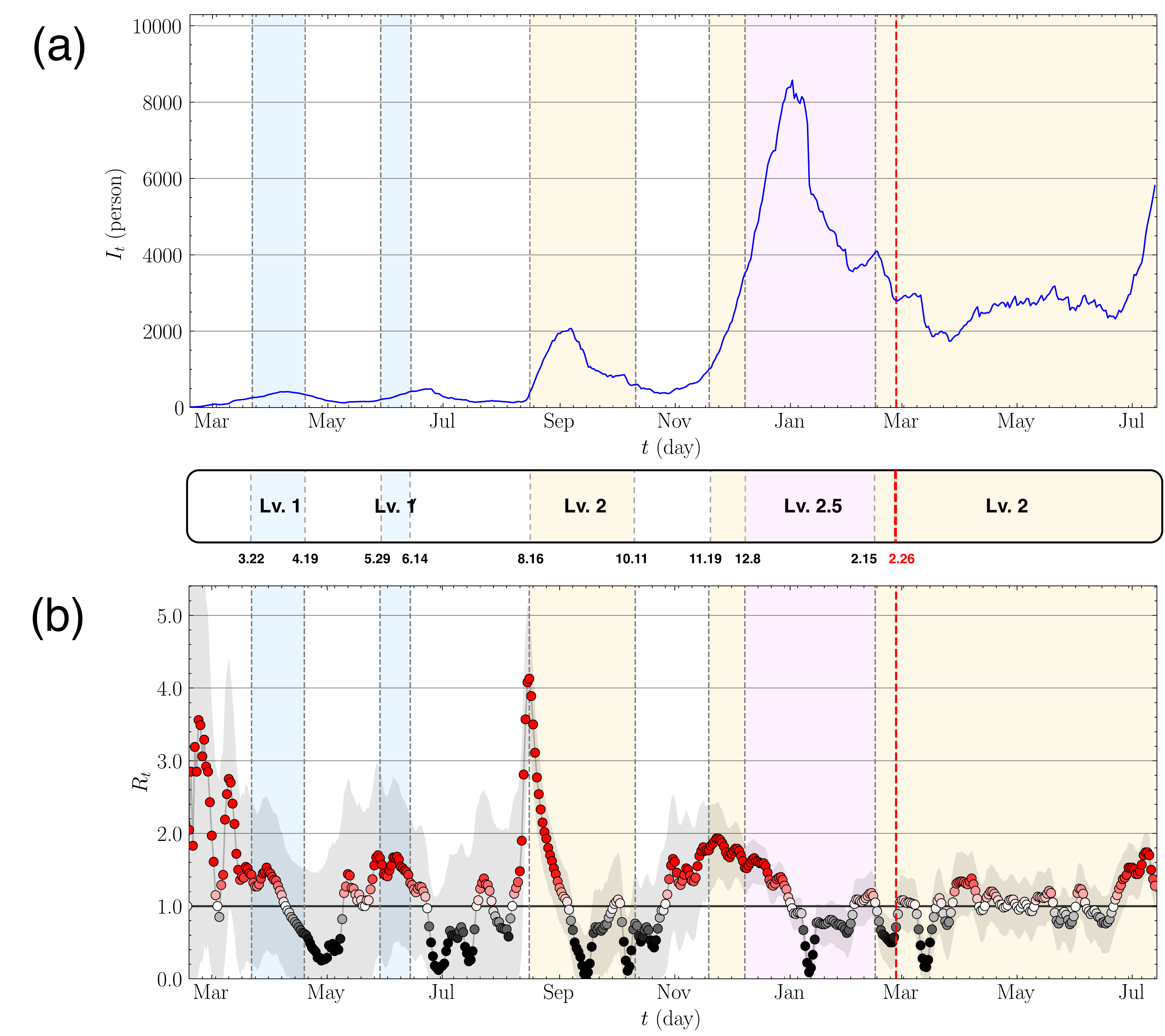}
\caption{Empirical data for the infected cases and the effective reproduction number in Seoul. (a) Time course of the infected cases is plotted by a solid line from Feb. 5, 2020 to Jul. 13, 2021 (blue). (b) The effective reproduction number $R_t$ is inferred by Bayesian inference in Eq.~(\ref{eq:bayes_theorem}). Representatively, we only plot the most probable $R_t$ [having maximum $P(R_t)$], so-called the mode of $R_t$ with color gradient corresponding to the value. The gray area enveloping the mode $R_t$ indicates the error bar defined as the $R_t$ in $0.05\leq P(R_t)\leq 0.95$ covering 90\% of $P(R_t)$. The mode of $R_t$ lies on around the center in the error bar. The horizontal solid line indicates $R_t=1$ for a guide to the eye. In both (a) and (b), the red dashed line indicates the first time to start vaccination in Korea. The colored regions highlight the periods to strengthen the social policy against pandemic. On the whole, the policies may work in curbing the spreading evidently from $R_t<1$. %\noteMJL{to GH: (1) change the axis labels as Date(day) $\to t$, Infected Population(person) $\to I_t$ (unify the label format in either words or notations). (2) removed all legends in (a) and "Start vaccination" in (b). (3) tone down the gray color. }.\noteGH{done!} \noteMJL{Checked.}
}
\label{fig:data_Rt}
\end{figure*}

%%% predetermined parameter value
%% virus-specific parameters
Since there are several parameters to describe our model, we would like to predetermine a few parameters from empirical data for reducing the number of parameters. The only virus-specific parameters are assumed as the inverse latent period $\sigma={{1}/{4.1 \,(\rm days)}} = 0.24$ (E$\to$I) and the recovery rates $\gamma$ (infrastructure) and $\gamma^{\prime}$ (natural healing). We assume $\gamma=\gamma^{\prime}={{1}/{12 \,(\rm days)}}=0.083$ for simplicity.
%% parameters inferred from empirical data 
To specify the rest, we concentrate on the epidemic outbreaks in Seoul from Feb. 5, 2020 to Feb. 25, 2021 [see Fig.~\ref{fig:data_Rt}(a)] with the total population $N=9.6\times10^6$, which is the capital of South Korea and a highly populated city. From the data, we infer the proportion of the asymptotic infected individual ($\rm{I_{A}}$) $p_A = 0.36$, and the quarantine rate $\delta={{1}/{4 \, (\rm days)}}=0.25$ (${\rm I_S}\to$Q).   The vaccination speed rate $\alpha$ ($\rm{S}\to\rm{V}$) is the inverse of the average time taken for individual vaccination. For example, the Korean vaccination made the plan to complete the vaccination for 70\% of people while 9 months, which corresponds to $\alpha=0.005$.

%% beta_S and beta_V
The remaining parameters are related to the transmission rate, $\beta_S (\rm{S}\to\rm{E})$ [leading to $\beta_V (\rm{V}\to\rm{E})$]. The  transmission rate is affected by social property as well as the biological feature of the virus. We try to evaluate the $\beta_S$ from empirical data by Bayesian inference, because the $\beta_S$ is related to the first step to enter the epidemic spreading process.
Except for $\beta_S$ and $\beta_V$, the predetermined values of the parameters are referred to the epidemiological investigation in Korea by the Korea Centers for Disease Control and Prevention, the National Statistical Office, and Refs.~\cite{SEIQR2, who}. 

As to connectivity, we have only taken account of a compartment model, that is, a well-mixed model (fully connected network in terms of network science). In contrast, individual contact and spreading path of disease in reality may be more heterogeneous, which gives us necessity for the epidemic spreading on complex networks, where exist super spreaders (hub nodes). In this work, however, we would like to put focus on building up our epidemic spreading model of Eq.~(\ref{eq:rate_eqs}) and analyzing the results of the dynamics, rather than considering the network topology.

%=========+=========+=========+=========+=========+=========+=========+=========
\section{The Effective Reproduction Number by the Bayesian Inference}
\label{sec:reproduction_number}

%%% effective reproduction number
The epidemic transmission rate $\beta_{S}$ is closely related to the basic reproduction number, which is the expected number of secondary infectees directly generated by one infected individual~\cite{NewmanBook}. The basic reproduction number at the early stage in the SIR model is well known as $R_0={{\beta}/{\gamma}}$ with $\beta$ being the transmission rate (${\rm S}\to{\rm I}$). The reproduction number along with the transmission rate not only is a biological constant for the pathogen but also plays a role of a criterion letting us clearly know the threshold to spread the disease: If $R_0>1$, the disease outbreak occurs over the population, and does not if $R_0<1$ (the case of $R_0=1$ is marginal). Herein we are concerned in the temporal change of the reproduction number at time $t$, i.e., the effective reproduction number $R_t$. Considering the parameters related to the infection period, we can write down the effective reproduction number as
\begin{equation}
\beta_{S, t} = R_t[(1-p_A)\delta+p_A\gamma].
\label{eq:beta}
\end{equation}
Equation~(\ref{eq:beta}) is acquired under the assumption of $R_t = {\beta_{S, t}\over (1-p_A)\delta + p_A \gamma}$ for the negligible $V$ (we have not handle the vaccination yet when estimating $R_t$), where the denominator $(1-p_A)\delta + p_A \gamma$ is interpreted as a total recovery rate in more general sense [note Eq.~(\ref{eq:rate_eq_i})].
%\noteMJL{to GH: What is the evidence for this formula? Our assumption?}
%\noteGH{It is our assumption according to the definition of $R_t = \frac{\beta_{s,t}}{\gamma_{total}}$ when $V$ is very small. we can modified $\gamma_{total}$ to $(1-p_A)\delta+p_A\gamma$} \noteMJL{Checked.} 
When all the infectees recover the disease without the quarantine ($\delta=0$ and $p_A=1$), Eq.~(\ref{eq:beta}) returns to the case of the SIR model as $\beta = \beta_{s, 0} = R_0\gamma$. The predetermined value $\delta$ in Sec.~\ref{sec:model} is used for estimating the $\beta_{S, t}$. After obtaining the $\beta_{S, t}$, we vary the $\delta$ later for comparing the suppression strategies.

%%% validity of the estimated R_t
\begin{figure*}
\includegraphics[width=\textwidth]{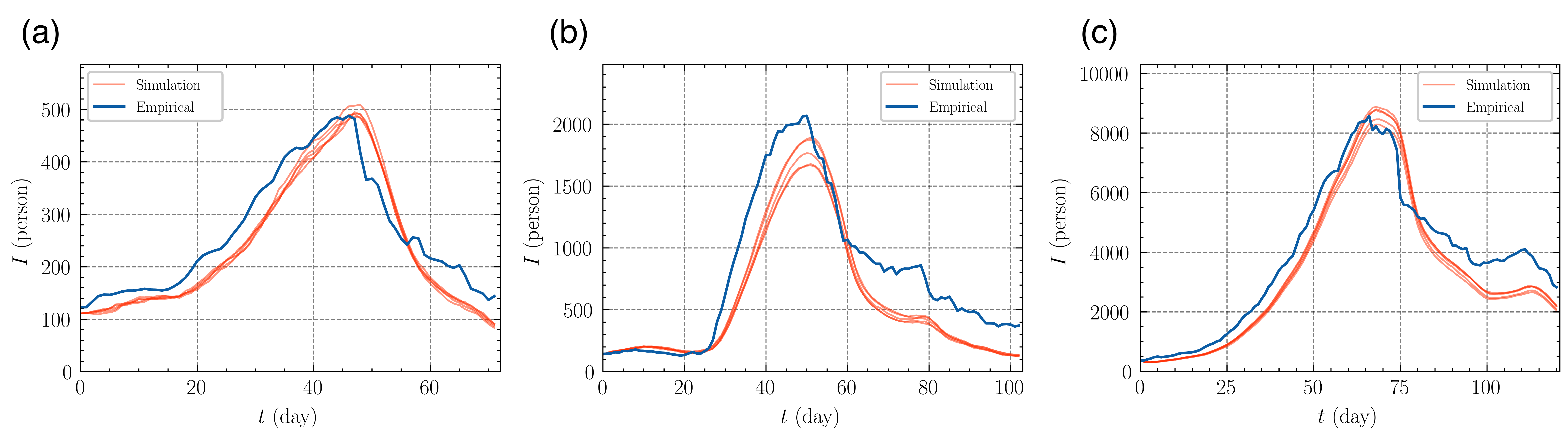}
\caption{The validity of $R_t$ by comparing $I_t$ in the empirical data and SVEIQR model. (a), (b), and (c) correspond to the 2nd, 3rd, 4th pandemic waves in Seoul, respectively. The starting time of each wave is set to be 0 on a horizontal axis. The blue line is the empirical data, and the red lines are the results of the numerical simulations. To implement the simulation, we use the mode of $R_t$ and set $\alpha=0$ not to consider the vaccination effect yet. %\noteMJL{A. using mode Rt, stochastic model. to GH:  Also, change the axes labels: Population(person) $\to I$, Simulation time(day) $\to t$.}\noteGH{done!}\noteMJL{Checked.}
}
\label{fig:simulation_I}
\end{figure*}

%%% Bayesian inference
To estimate the epidemic transmission rate $\beta_{S, t}$ from empirical data via the effective reproduction number $R_t$ in Eq.~(\ref{eq:beta}), we employ the Bayesian inference represented as
\begin{equation}
    P(R_t|I_t) = \frac{P(I_t|R_t)P(R_t)}{P(I_t)},
\label{eq:bayes_theorem}
\end{equation}
where $I_t$ is the number of the infected cases at time $t$. In the empirical COVID-19 data in Seoul, the confirmed cases are recorded instead of the infected cases, but we regard the infected people at time $t$, $I_t$ as the number of the infected person at time $t$, i.e., the cumulative confirmed cases until time $t$ minus recovered cases at time $t$ (it is impossible to keep track of the infected-but-not-confirmed cases).  
Taking a case that the infection event independently occurs for given $R_t$, then the $P(I_t|R_t)$ follows the Poisson distribution from the Poisson process:
\begin{equation}
    P(I_t|R_t) = \frac{{\lambda_t} ^{I_t} e^{-\lambda_t}}{I_t!},
\label{eq:likelihood}
\end{equation}
with a mean value $\lambda_t \equiv I_{t-1}e^{\gamma(R_t-1)}$. An assumption on the mean value $\lambda_t$ of $I_t$ is that it is increased by $\beta_t I_{t-1}$ and decreased by $\gamma I_{t-1}$, that is, $\lambda_t \simeq I_t + (\beta_t-\gamma)I_t\simeq I_t e^{\gamma(R_t-1)}$. For the calculation of $R_t$, the simple SIR process are considered and the basic reproduction number from time $t-1$ to $t$ represents $R_t={{\beta_t}/{\gamma}}$. %\noteMJL{to SW: can you describe the validity of the assumption (I mean the reason why we assume the SIR model for evaluating the Rt)? By the way, do we have to argue this? if unnecessary, we can remove the additional explanation.}

%%% how to estimate the likelihood
The likelihood $P(I_t|R_t)$ and the prior probability $P(R_t)$ can be iteratively estimated. First, for simplicity, the prior probability $P(R_0)$ at $t=0$ follows the uniform distribution ranging from 0 to 10. Based on a hypothesis that $P(I_t|R_t)$ and $P(I_{t+1}|R_{t+1})$ do not differ so much, we estimate the $R_{t+1}$ for a given $R_t$ with the conditional probability $P(R_{t+1}|R_t)$ which follows the Gaussian distribution with the mean $R_t$ and we add the heuristically-determined small standard deviation 0.1.  %\noteMJL{to GH: what is the reference of this method? Especially the EQ.(4) with the mean and the Gaussian noise. What range does the reference(PLoS One 3, e2185) cover in this method?}.
%\noteGH{PLoS One paper cover that method. Assuming that $P(I_t|R_t)$ and $P(I_{t+1}|R_{t+1})$ will not be much different, we multiplied $P(I_t|R_t)$ by Gaussian noise to make $P(I_{t+1}|R_{t+1})$.} \noteMJL{Checked.}

Equipping with the empirically estimated $\beta_S$ as the realistic baseline, we compare the two vaccination and quarantine strategies to suppress the spreading. The vaccination is the strategy to focus on inhibiting the onset of infected cases, and the quarantine is on doing the spreading of already infected individuals. We vary the values of the corresponding parameters $\alpha$ and $\delta$ and investigate how much reduced the infected cases by simulating the SVEIQR model in Eq.~(\ref{eq:rate_eqs}).

%=========+=========+=========+=========+=========+=========+=========+=========

\section{Results}
\label{sec:results}

%%% short explanation about I_t in Seoul
To utilize the plausible epidemic transmission rate $\beta_{S, t}$ as the test bed for applying suppression strategies, we first estimate the effective reproduction number $R_t$ based on Eq.~(\ref{eq:beta}), where $I_t$ is extracted from the empirical data plotted in Fig.~\ref{fig:data_Rt}. For convenience's sake, we distinguish the pandemic by clarifying four different waves by using the local minima~\footnote{the 1st wave from Feb. 5, 2022 to May 9, 2022, the 2nd wave from May 10, 2022 to Jul. 19, 2021, the 3rd wave from Jul. 20, 2021 to Oct. 28, 2021, and the 4th wave from Oct. 29, 2021 to Feb. 25, 2022}. Differently from the other waves, we end the 4th wave forcibly right before starting vaccination (the red dashed line in Fig.~\ref{fig:data_Rt}). Thus, we only cover the uncontaminated data in terms of the absence of vaccination. Whenever perceiving an onset of a notable wave, the Korean government enforces the spreading-suppression policy such as the social distancing and quarantine. One sees that the social policies may work somewhat because the waves temporarily shrank right after the enforcement. 

%%% estimation of the R_t
The effective reproduction number $R_t$ in Fig.~(\ref{fig:data_Rt}) more clearly demonstrates the policy effect. We almost always observe the unimodal distribution of $R_t$ obtained by the method in Sec.~\ref{sec:reproduction_number}, so choose the most probable $R_t$, so-called the mode in statistics, as a representative value. The mode of $R_t$ is plotted in Fig.~\ref{fig:data_Rt}(b) with enveloped by the range between 5- and 95-percentile, of which narrowness along with its central position within the range can be the evidence of the mode's representability . During the enforcement of the social policy, the mode value of $R_t$ tends to be less than 1 ($R_t<1$). When taking that the social policy does not immediately work for the shrinkage of the spreading, the time-delayed behavior of the estimated $R_t$ seems understandable. The tendency makes us believe the inference process in Sec.~\ref{sec:reproduction_number} to be trustworthy, so we use the inferred value of the mode $R_t$ to evaluate the $\beta_{S, t}$ governing Eq.~(\ref{eq:beta}).

%%% vaccination and quarantine
\begin{figure*}
\includegraphics[width=\textwidth]{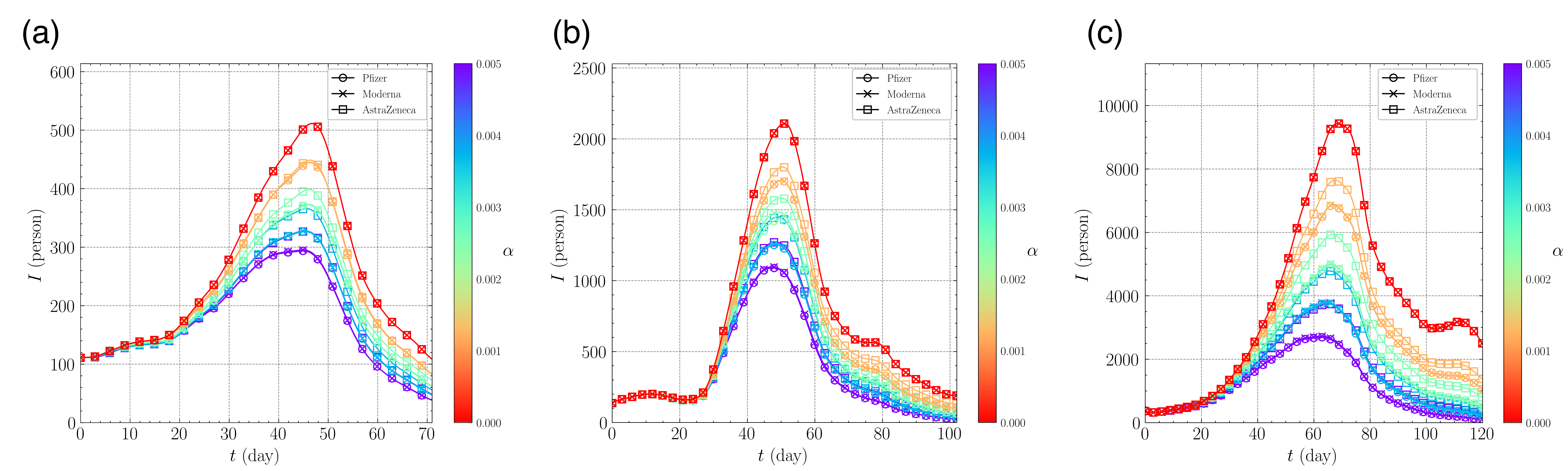}
\caption{The vaccination effects in terms of the vaccination speed and the efficacy in the three pandemic waves in Seoul for (a) the 2nd, (b) 3rd, and (c) 4th wave. The color indicates the vaccination speed, and the efficacy of vaccine is distinguished by symbol as open circle for the Pfizer ($\epsilon=0.95$), cross for the Moderna ($\epsilon=0.94$), and open square for the AstraZeneca ($\epsilon=0.7$). Every curve is obtained by averaging the simulation results with applying the given $\alpha$ to the $I_t$ curve labeled ``Simulation'' in Fig.~\ref{fig:simulation_I}. %\noteMJL{Why don't these curves fluctuate? Are they the averaged curves or did you use the representative Rt? Also, change the axes labels: Population(person) $\to I$, Simulation time(day) $\to t$. Change the legend as Pfizer vaccinated $\to$ Pfizer, Moderna vaccinated $\to$ Moderna, Astrazeneca vaccinated $\to$ AstraZeneca.}\noteGH{done!}\noteMJL{Checked.}
}
\label{fig:vaccination}
\end{figure*}

Before applying the level of suppression strategy (vaccination and quarantine speed) characterized by the relevant parameters ($\alpha$ and $\delta$) to our SVEIQR model in Eq.~(\ref{eq:rate_eqs}), we confirm the validity of the mode $R_t$ or equivalently $\beta_{S,t}$ by comparing the $I$ curves obtained from the simulation with the predetermined parameters in Sec.~\ref{sec:model} to the $I$ from the empirical data. As mentioned, we only handle the empirical $I_t$ without political adoption of vaccination (right before the red dashed line in Fig.~\ref{fig:data_Rt} or until Feb. 25, 2022). Hence, in the simulation as well, we set $\alpha=0$ after taking the respective initial $I$ from every wave, leading to the maintenance of $V=0$. One sees the comparison in Fig.~\ref{fig:simulation_I}. Due to the reliability of the mode $R_t$ at the initial stage [see the relatively large error bar in Fig.~\ref{fig:data_Rt}(b)], we exclude the 1st wave in analysis throughout this paper. The numerical integration of Eq.~(\ref{eq:rate_eqs}) is also implemented and its result not shown here is in accordance with that of the simulation in Fig.~\ref{fig:simulation_I}.

%%% focus on vaccination strategy and introduce the efficacy value  
Now we investigate the effect of the suppression strategy in the viewpoint of the social policy: the enforcement of the vaccination and quarantine whose speeds are parameterized by $\alpha$ and $\delta$, respectively. While controlling $\alpha$ and $\delta$, the efficacy of the vaccine affects the suppression effect indirectly. The efficacy $\epsilon$ weakens the possibility that the vaccinated individual turns into the infected with $\beta_V=(1-\epsilon)\beta_S$ in this model. There are three representative companies which produce vaccines for the COVID-19, Pfizer, Moderna, and AstraZeneca. The efficacy with the confidence level (CI) of their products are 95\% (CI: 90.0\% -- 98.0\%), 94\% (CI: 88.0\% -- 98.0\%), and 70\% (CI: 56.0\% -- 85.0\%), respectively, which are clinically obtained.

%%% results of the vaccination
Figure~\ref{fig:vaccination} shows the number of infected individuals depending on the three different efficacy $\epsilon$ (distinguished by symbol) for a given vaccination speed $\alpha$ (represented by color) with the inferred $\beta_{S, t}$ in Sec.~\ref{sec:reproduction_number} in each wave. We increase the value of $\alpha$ up to $0.005$, which corresponds to the goal of the vaccination policy in Korea to fully vaccinate about 70\% of Korean over nine months. 
When using the Pfizer vaccine (symbolized by open circle) with the highest efficacy among the three, the height of the peak under the plan ($\alpha=0.005$, colored by purple) are maximally decreased by 42\%, 48\%, and 71\% in three waves, respectively, compared to the case of no vaccination ($\alpha=0$, colored by red). The AstraZeneca vaccine with the lowest efficacy shows the minimal decrease of the height of the peak in the three waves as 34\%, 40\%, and 60\%, respectively. The suppression effect is lower than that of the Pfizer vaccine, but overlaps of some curves composed of the combinations of different $\alpha$ (color) and types of vaccine (symbol) in Fig.~\ref{fig:vaccination} tells us that increasing the vaccination speed with the low-efficacy vaccine similarly works to using the high-efficacy vaccine at the lower vaccination speed. For instance, in Fig.~\ref{fig:vaccination}(a), the curve when using the Pfizer vaccine ($\epsilon=0.95$) at the speed $\alpha=0.0004$ (blue circle) is almost the same as the curve when using the AstraZeneca vaccine ($\epsilon=0.7$) at $\alpha=0.0005$ (purple square), which means the two combinations of the efficacy and the speed have almost the same effect.

\begin{figure}[t]
\includegraphics[width=1.1\columnwidth]{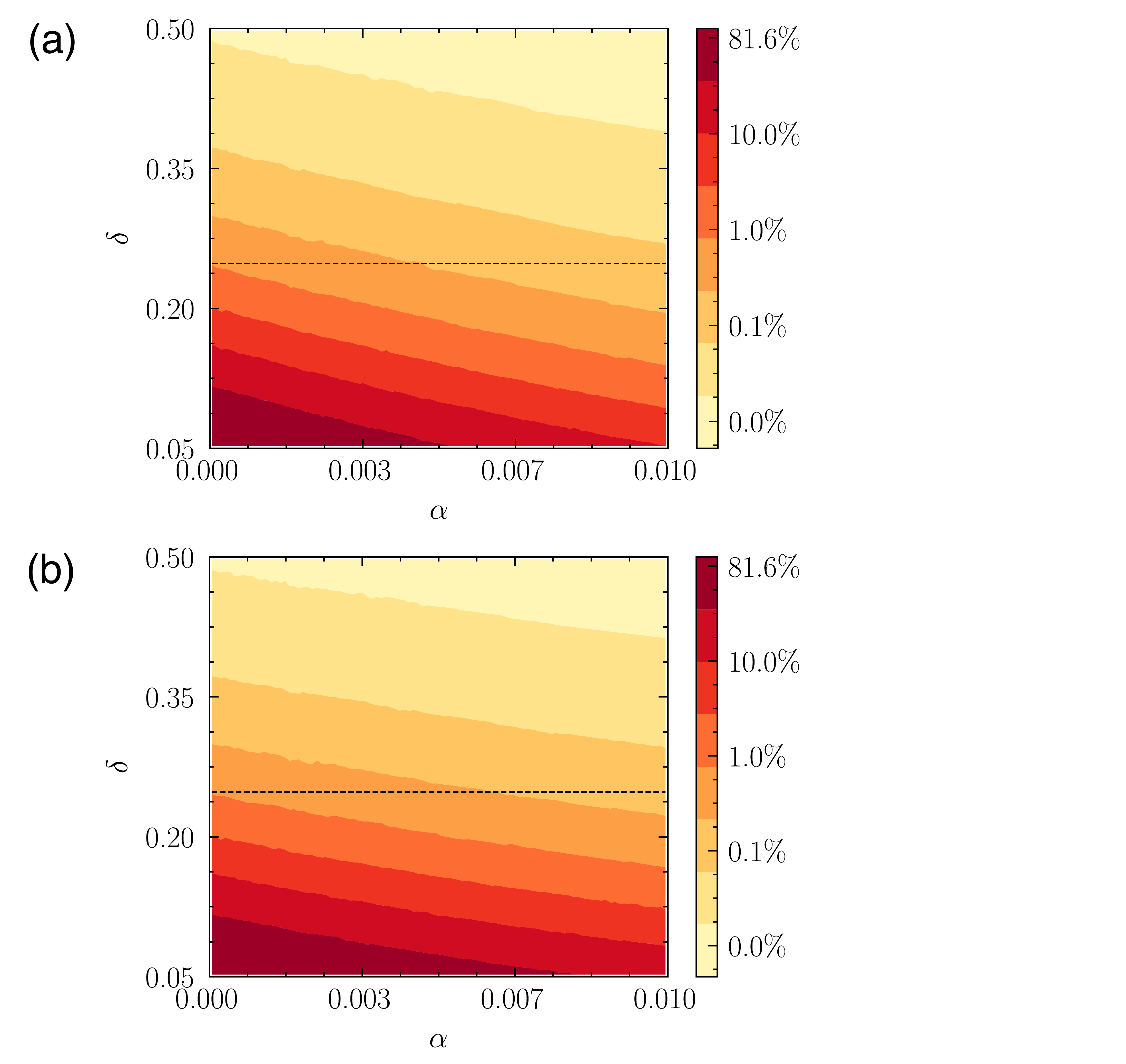}
\caption{The relative final size of the outbreak on $\alpha$-$\delta$ plane with $\alpha$ and $\delta$ being the vaccination and quarantine speed, respectively, for given vaccines such as (a) Pfizer and (b) AstraZeneca. The color bar depicts the relative size which is evaluated by the governing equation of the SVEIQR model when applied to the 4th waves in Fig.~\ref{fig:simulation_I}(c). The dashed horizontal line indicates the empirical value $\delta={1 \over {\rm 4 days}}=0.25$ corresponding to the in-force quarantine speed at the time. %\noteGH{The total fraction of the infected individuals on $\alpha$-$\delta$ plane with $\alpha$ and $\delta$ being the vaccination and quarantine speed, respectively, for given vaccines such as (a) Pfizer (b) AstraZeneca. The dark black line means quarantine speed applied in Korea, $\delta=0.25$. I prepared two versions of the figure that are used "imshow" and "contour" figure. So please check which one is better between "result\_comparing" and "result\_comparing\_discrete"}\noteMJL{Checked. I placed two versions for comparison. Which one is used will be decided by SW!}
}
\label{fig:alpha_delta}
\end{figure}

%\noteMJL{to GH: Why the value of the delta is so small, compared with the predetermined value 0.25?}\noteGH{Due to the large difference in the influence of alpha and delta, the pattern of total confirmed patients can be confirmed only by matching a similar order, so the delta was reduced and confirmed. However, as a result of the discussion, the scope was changed and the simulation was performed again. }\noteMJL{Checked.}
Now, we vary the vaccination and quarantine speeds simultaneously around the empirical value of $\alpha=0.005$ and $\delta=0.25$, using the Pfizer ($\epsilon=0.95$) and AstraZeneca ($\epsilon=0.7$) vaccines. We representatively apply to the 4th pandemic waves [labeled as ``Simulation'' in Fig.~\ref{fig:simulation_I}(c)]. %\noteMJL{to GH: please check whether it's correct.}] \noteGH{exactly collect!}
Figure~\ref{fig:alpha_delta} exhibits the relative outbreak size computed as ${{I+Q+R}\over N}$ at the final time [refer to the color bar]. In the vaccine-wise comparison, one sees the different sloped of $\delta$ with respect to $\alpha$ depending on efficacy $\epsilon$. Of course, we need to take care to give meaning to the slope value itself because of the different scales of $\alpha$ and $\delta$ resulting from the fact that the vaccination and quarantine are inherently different strategies. From the steepness of the slope, we catch how much the suppression effect is sensitive to adjustment of the strategies. The result says that the higher-efficacy vaccine [Figs.~\ref{fig:alpha_delta}(a)] lowers the outbreak size faster than the lower-efficacy one [Figs.~\ref{fig:alpha_delta}(b)] for a given quarantine speed $\delta$. In other words, it also means that, if we only have access to the low-efficacy vaccine, boosting the vaccination speed up (increasing $\alpha$) shows the similar effect of outbreak suppression to the low vaccination speed with the high-efficacy vaccine. In addition, the in-force quarantine speed ($\delta=0.25$ represented by a dashed line) play an quite important role: For $\alpha=0$ at the time of the 4th wave, if the infected individual had been found in a day or two late (leading to $\delta={1 /{\rm 5 (days)}}=0.2$ and $\delta={1 / {\rm 6 (days)}}\simeq 0.17$), the outbreak size would have grown out of control. The qualitatively similar behavior is observed in all $\alpha$'s. 

%%% vaccination and quarantine together 
%=========+=========+=========+=========+=========+=========+=========+=========

\section{Summary and Discussion}
%%% summary
To explore the effects of the quarantine and vaccination as the representative policy to hinder the epidemic spreading, we have proposed a susceptible-vaccinated-exposed-infected-quarantined-recovered (SVEIQR) model of Eq.~(\ref{eq:rate_eqs}). To treat the recent pandemics, i.e., COVID-19, we have utilized the clinical (for vaccine efficacy) and empirical data (for the time varying reproduction number) and experimented with the suppression strategies applying to the infected cases in Seoul. To reflect the empirical cases more concretely, we obtained the reproduction number based on the abundance of the infected cases by the Bayesian inference. After the inference, we have probed the effect of vaccination and quarantine speeds on the suppression of the infected cases with considering the three different types of vaccine. As shown in Fig.~\ref{fig:alpha_delta}, one can find the possible suppression level for given vaccination and quarantine speeds ($\alpha$ and $\delta$) and capture the two features: (i) Regarding the higher-efficacy vaccine, the vaccination speed $\alpha$ is more sensitive to a decrease in the outbreak; 
(ii) It would be likely to drastically increase the outbreak size if a quarantine speed was slower than the current policy. The result can be utilized properly in such a way that one determines how fast speeds need to achieve a desirable suppression level depending on the vaccine efficacy. Now we adopted several types of vaccines instead of one type solely, so can check an effect of the mixed ratio of vaccines (with different efficacy). Policymakers can refer to the quantitative approach to make decisions for public health.

As mentioned before, in terms of connectivity, we have only dealt with the fully connected cases, not considering complex temporal social/contact networks. Exploring this model on heterogeneous networks makes us do enriched analyses, such as whether the $k$-core still determines the most effective super-blocker which is the well-known result in the previous research~\cite{RN55}, what node properties mainly correlated with the suppression effect under this mechanism, and so on. In addition, one can refine more the parameters in the proposed SVEIQR model with taking account of demography. Even if exposed to the same pathogen, the fatality of the pathogen can be up to the biological condition of the exposed individual. Besides, the vigorous social activity can be a major factor of the spreading. The both will presumably affect an outbreak size, which is worthy to be considered. They share a common feature to characterized by age. The biological condition of human body can depend on age, and the contact (including social interaction) rate between people representing the vigorous social activity can be different from age to age. The vaccination strategy can be also modified depending on age, such as the vaccine type (with different efficacy), priority of vaccination, vaccination speed, and so on. To deal with in our model framework, we need to modify the relevant parameters using the demographic information, e.g., the age distribution, and the contact rate between ages. Together with complex connectivity, we leave this argument being a future work.

%=========+=========+=========+=========+=========+=========+=========+=========

\begin{acknowledgments}
This research was supported by the National Research Foundation (NRF) of Korea through the Grant Nos. NRF-2020R1A2C2010875 (S.-W.S.), NRF-2021R1C1C1007918 (M.J.L.). This work was also partly supported by Institute of Information \& communications Technology Planning \& Evaluation (IITP) grant, funded by the Korea government (MSIT), No.2020-0-01343, Artificial Intelligence Convergence Research Center (Hanyang University ERICA) (S.-W.S.). We also acknowledge the hospitality at APCTP where part of this work was done.
\end{acknowledgments}

%=========+=========+=========+=========+=========+=========+=========+=========
%\noteGH{There are no strict requirements on reference formatting at submission. The reference style used by the journal will be applied to the accepted article by Elsevier at the proof stage. So I used the bibtex file provided} \noteMJL{Checked.}
\bibliographystyle{elsarticle-num}
\bibliography{reference}

%=========+=========+=========+=========+=========+=========+=========+=========

\end{document}